\documentstyle{article}
\newcommand {\ip}[1]{\left\lfloor #1\right\rfloor}
\newcommand {\figsz}{\protect\footnotesize}
\newcommand {\tsub}[1]{_{\mbox{\protect\scriptsize #1}}}
\newcommand {\tsup}[1]{^{\mbox{\protect\scriptsize #1}}}

\newcommand {\Ref}[1]{(\ref{#1})}
\newcommand
{\Stretch}[1]{\renewcommand{\baselinestretch}{#1}\large\normalsize}
\newcommand {\figureStretch}{\Stretch{0.9}}
\newcommand {\footnoteStretch}{\Stretch{1}}
\newcommand {\unStretch}{\Stretch{1}}
\newcommand {\arrayStretch}{0.3em}
\begin{document}

\pagestyle{empty}

\Stretch{3}
\begin{center}
{\huge \bf One Dimensional Continuum
Falicov-Kimball Model in the Strongly Correlated Limit}
\vspace*{5em}

\Stretch{1.5}
{\large
Robert James Bursill
\vspace*{2em}

Department of Mathematics\\
The University of Melbourne
\\Email: r.bursill@sheffield.ac.uk}
\vspace*{2em}

Short Title -- Continuum Falicov-Kimball Model\\
Classification Number -- 05.30 -- Statistical Mechanics---Quantum
\end{center}

\vfill
\eject

\Stretch{1}

\begin{abstract}
In this paper we study the thermodynamics of the one dimensional continuum
analogue of the Falicov-Kimball model in the strongly correlated limit
using a method developed by Salsburg, Zwanzig and Kirkwood for the Takahashi
gas. In the ground state it is found that the $f$ electrons form
a cluster. The effect of including a Takahashi repulsion between
$f$ particles is also studied where it is found that as the repulsion is
increased the ground state $f$ electron configuration changes
discontinuously from the clustered configuration to a homogeneous or equal
spaced configuration analogous to the checkerboard configuration which arises
in the lattice Falicov-Kimball model.
\end{abstract}

\setcounter{page}{0}
\vfill
\eject

\pagestyle{plain}

\section{Introduction}
\label{Introduction_Falicov_Kimball}
\setcounter{equation}{0}

The Falicov-Kimball model \cite{Falicov_Kimball} was initially proposed as a
simple model for the semiconductor-metal
transition in $\mbox{SmB}_{6}$ and transition metal oxides. The model
consists of two species of electrons---localised
$f$ electrons and a itinerant band of $d$ electrons. The Hamiltonian is
\begin{equation}
{\cal H}=\sum_{k}\epsilon_{k}c_{k}^{\dagger}c_{k}
+U\sum_{i}d_{i}^{\dagger}d_{i}f_{i}^{\dagger}f_{i}
\label{HFK}
\end{equation}
where $f_{i}$ is the destruction operator for $f$ electrons on site $i$,
$d_{i}$ is the destruction operator for the $d$-band
Wannier state at site $i$ and $c_{k}$ is the destruction operator for the $d$
electron Bloch state with wave vector $k$ and
kinetic energy $\epsilon_{k}$.

The only interaction then is an on-site interaction between $f$ and $d$
electrons. Of course the model could be extended to
include longer range $f$-$d$ interactions, $f$-$f$ (lattice gas type)
interactions and $d$-$d$ interactions. Spin could be
introduced for the $d$ electrons with Hubbard-type interactions. In
\cite{Kennedy_Lieb} and \cite{Lieb} the model was
independently introduced as a model for crystallization and also
viewed as a Hubbard model \cite{Hubbard} where one of the spin species of
electrons is immobile.

A great deal more is known about the Falicov-Kimball model than is known
about the Hubbard model. In
\cite{Falicov_Kimball} the model was treated within a mean field framework.
It was deduced that for large enough $U$ there
is a first order phase transition which was interpreted as a
semiconductor-metal transition.

In \cite{Kennedy_Lieb} it was established by means of operator
inequalities that for the bipartite lattices with equal numbers of $f$ and
$d$ electrons
and a half-filled band, the ground state is ordered in any dimension, the $f$
electrons
forming a checkerboard pattern. For
$d\geq 2$ it was shown further that the crystalline order persists at finite
temperature.

In \cite{Brandt_Schmidt1} and \cite{Brandt_Schmidt2} sharp upper and lower
bounds are obtained for the ground state
energy in the two dimensional case. For some special values of the parameters
the bounds are shown to
coalesce and so the exact ground state energy is derived. Some of the results
of \cite{Kennedy_Lieb} and \cite{Lieb} are also
independently reproduced. In \cite{Brandt_Mielsch1} and
\cite{Brandt_Mielsch2} the thermodynamics of the model is
investigated for large dimensionality $d$. An exact solution is obtained in
the $d=\infty$ limit for the thermodynamic
potential, the critical
temperature, the order parameter and several correlation functions.

In \cite{Freericks} the ground state of the one dimensional model is studied
by means of second order perturbation theory and
exact numerical calculations for periodic configurations of the $f$
electrons. A restricted ground state phase diagram with a
fractal structure is derived. In \cite{Gruber} the one dimensional model is
studied for large negative $U$, the $f$ electrons
being interpreted as {\em nuclei}. An exact formula for the leading order
contribution (for large $|U|$) to the ground state
energy is derived from which it is deduced that the system forms {\em atoms}
($d$-$f$ pairs) which have an effective
repulsion between them if $|U|$ is sufficiently large.

In this paper we study the one dimensional continuum
analogue of the Falicov-Kimball model in the strongly correlated limit.
Although in terms of mathematical complexity this
model is a long way removed from the lattice model with finite interaction
strength in higher dimensions, there may be some
generally applicable
physics which can be deduced from this study.
We obtain the exact thermodynamic potential via a method first
used by Salsburg, Zwanzig and Kirkwood \cite{SZK} for the Takahashi gas
\cite{Takahashi}. We discuss the ground state energy as well as a physical
interpretation of the model in terms of the Takahashi gas. Finally, we
discuss
the effect of introducing a Takahashi type
interaction between the $f$ particles.

\section{Continuum Model and Exact Solution in One Dimension in the Limit of
Strong Correlations}
\label{Continuum_Model}
\setcounter{equation}{0}

In the continuum analogue of the Falicov-Kimball model, we have a system of
fermions ($d$ electrons) and stationary
scatterers ($f$ electrons). If there are $M$ scatterers at positions
$y_{1},\ldots,y_{M}$ then the equation for the
$d$ particle eigenfunction $\psi(x)$ is
\begin{equation}
\left[-\frac{1}{\pi^{2}}\nabla^{2}+U\sum_{j=1}^{M}\delta\left(x-
y_{j}\right)\right]\psi(x)=E\psi(x)
\label{Eigenvalue_Problem}
\end{equation}
where $E$ is the energy eigenvalue and we have scaled the fermion mass and
Planck's constant appropriately.

In the strongly correlated limit $(U=\infty)$ the one dimensional continuum
analogue of the Falicov-Kimball model reduces
to a system containing a number of impenetrable barriers configured along the
line. That is, if the length of the line is $V$ and
there are $M$ $f$ electrons at positions $0<y_{1}<y_{2}<\ldots<y_{M}<V$ then
the $d$ electrons experience impenetrable
barriers at positions $y_{0}\equiv 0,y_{1},\ldots,y_{M}$, and
$y_{M+1}\equiv V$\footnoteStretch\footnote{We assume fixed end boundary
conditions.}\unStretch~or infinitely high wells in the regions
$\left(y_{0},y_{1}\right)$,
$\left(y_{1},y_{2}\right)$,\ldots,$\left(y_{M},y_{M+1}\right)$.

In this case the eigenvalue problem \Ref{Eigenvalue_Problem} is trivial,
reducing to
\begin{eqnarray}
-\frac{1}{\pi^{2}}\psi''(x) & = & E\psi(x)
\\[0.3em]
\psi(y_{j}) & = & 0\;\mbox{ for }\;j=0,1,\ldots,M+1
\\[0.3em]
\psi(x) & = & 0\;\mbox{ for }\;x<0\mbox{ and }x>V
\end{eqnarray}
The single particle wavefunction for a state with wave number $k\geq 1$ in
well $1\leq j\leq M+1$ is
\begin{equation}
\psi_{jk}(x)=\left\{
\begin{array}{ll}
\sqrt{\frac{2}{y_{j}-y_{j-1}}}\sin\left(
\frac{\pi k\left(x-y_{j-1}\right)}{y_{j}-y_{j-1}}\right) &
y_{j-1}\leq x\leq y_{j}
\\[\arrayStretch]
0 & \mbox{otherwise}
\end{array}
\right.
\end{equation}
and the corresponding energy eigenvalue is
\begin{equation}
E_{jk}=\frac{k^{2}}{\left(y_{j}-y_{j-1}\right)^{2}}
\label{Ejk}
\end{equation}

To calculate the thermodynamic potential for this model, we begin by defining
a canonical partition function for the system with $M$ $f$ electrons and $N$
$d$ electrons viz
\begin{equation}
Z_{MN}\equiv\int{\cal D}y\prod_{j=1}^{M+1}\prod_{k=1}^{\infty}
\sum_{n_{jk}=0,1}'\exp\left(-\beta n_{jk}E_{jk}\right)
\label{Partition_Function}
\end{equation}
where the primed sum is restricted to configurations with $N$ $d$ particles
ie
\begin{equation}
\sum_{j=1}^{M+1}\sum_{k=1}^{\infty}n_{jk}=N
\end{equation}
and
\begin{equation}
\int{\cal D}y\equiv\int_{0}^{V}dy_{M}
\int_{0}^{y_{M}}dy_{M-1}\ldots\int_{0}^{y_{2}}dy_{1}
\label{intDy}
\end{equation}

The restriction on the sum can be removed by defining a grand canonical
partition function
\begin{eqnarray}
{\cal Q}(V) & \equiv & \sum_{N=1}^{\infty}z^{N}Z_{MN}
\\[0.3em]
& = & \int{\cal D}y\prod_{j=1}^{M+1}\prod_{k=1}^{\infty}\left[
1+ze^{-\beta E_{jk}}\right]
\label{Q=FK}
\end{eqnarray}
where use has been made of \Ref{Partition_Function}.

Making the definition
\begin{equation}
f(x)\equiv\prod_{k=1}^{\infty}\left[1+ze^{-\beta k^{2}/x^{2}}\right]
\label{f}
\end{equation}
we have, using \Ref{Ejk}, \Ref{intDy}, \Ref{Q=FK} and \Ref{f} that
\begin{eqnarray}
{\cal Q}(V)
& = &
\int{\cal D}y\prod_{j=1}^{M+1}
f\left(y_{j}-y_{j-1}\right)
\\[0.3em]
& = &
\int_{0}^{V}
f\left(V-y_{M}\right)dy_{M}
\int_{0}^{y_{M}}f\left(y_{M}-y_{M-1}\right)dy_{M-1}
\nonumber
\\[0.3em]
& & \hspace{15em}\ldots\int_{0}^{y_{2}}
f\left(y_{2}-y_{1}\right)
f\left(y_{1}\right)dy_{1}
\label{Grand_Canonical_Partition_Function}
\\[0.3em]
& = &
(f\ast\ldots\ast f)(V)
\label{Q==FK}
\end{eqnarray}
where $f\ast\ldots\ast f$ denotes the $(M+1)$-fold Laplace convolution of
$f$.

We take the Laplace transform of ${\cal Q}$ with respect to V viz
\begin{eqnarray}
{\cal L}_{{\cal Q}}(s)
& \equiv &
\int_{0}^{\infty}e^{-sV}{\cal Q}(V)dV
\\[0.3em]
& = &
\int_{0}^{\infty}e^{-sV}
(f\ast\ldots\ast f)(V)
\\[0.3em]
& = &
\left[F(s)\right]^{M+1}
\label{LQ=}
\end{eqnarray}
where
\begin{equation}
F(s)\equiv\int_{0}^{\infty}f(x)e^{-sx}dx
\label{Fcunt}
\end{equation}
denotes the Laplace transform of $f$ and we have made use of \Ref{Q==FK} and
the Laplace
convolution theorem.

Applying the Laplace inversion formula to \Ref{LQ=} we then have
\begin{equation}
{\cal Q}(V)=
\frac{1}{2\pi i}\int_{c-i\infty}^{c+i\infty}
\left[F(s)\right]^{M+1}e^{sV}ds
\label{Q===FK}
\end{equation}
where the contour lies to the right of all singularities of $F(s)$.

The singularity structure of $F$ can be deduced from a different
representation of $f$ which we derive in
appendix~\ref{Infinite_Product} viz
\begin{equation}
f(x)=\frac{e^{\chi\tsub{f}x}}{\sqrt{1+z}}
\prod_{k=1}^{\infty}
\left|1-e^{2\pi\left(-v_{k}+iu_{k}\right)x}\right|^{2}
\label{Large_x_Representation}
\end{equation}
where
\begin{equation}
\chi\tsub{f}\equiv\int_{0}^{\infty}\log
\left(1+ze^{-\beta x^{2}}\right)dx
\label{chif}
\end{equation}
denotes the thermodynamic potential for a system of free $d$ particles,
\begin{eqnarray}
u_{k} & \equiv &
\sqrt{\frac{
\mu+\sqrt{\mu^{2}+(2k-1)^{2}\pi^{2}/\beta^{2}}}{2}}
\label{uk}
\\[0.3em]
v_{k} & \equiv &
\frac{(2k-1)\pi}
{\sqrt{2}\beta\sqrt{\mu+\sqrt{\mu^{2}+(2k-1)^{2}\pi^{2}/\beta^{2}}}}
\label{vk}
\end{eqnarray}
The $u_{k}$ and $v_{k}$ form increasing sequences so $F(s)$ has simple poles
at
\begin{equation}
\left.
\begin{array}{rcl}
s & = & \chi\tsub{f} \\[\arrayStretch]
s & = & \chi\tsub{f}-2\pi\left(v_{k}\pm iu_{k}\right)\;\;k=1,2,3,\ldots
\\[\arrayStretch]
s & = & \chi\tsub{f}-4\pi v_{k}\;\;k=1,2,3,\ldots \\[\arrayStretch]
s & = & \chi\tsub{f}-2\pi\left(v_{k}\pm iu_{k}+v_{l}\pm
iu_{l}\right)\;\;1\leq k<l
\\[\arrayStretch]
& \vdots &
\end{array}
\right\}
\label{poles}
\end{equation}
In \Ref{Q===FK} then, we must take $c>\chi\tsub{f}$.

Now the thermodynamic potential $\chi$ is given by
\begin{equation}
\chi\equiv\lim_{V\rightarrow\infty}\frac{\log{\cal Q}(V)}{V}
\label{chiFK}
\end{equation}
where in taking the limit, the well density
\begin{equation}
m\equiv\frac{M+1}{V}
\end{equation}
is held constant. We evaluate the right hand side of \Ref{Q===FK} for large
$V$ and $M$ by the method of steepest
descents. We write
\begin{equation}
{\cal Q}(V)=\frac{1}{2\pi i}\int_{c-i\infty}^{c+i\infty}
\exp\left[VH(s)\right]ds\hspace{2em}c>\chi\tsub{f}
\end{equation}
where
\begin{equation}
H(s)\equiv s+m\log F(s)
\label{H(s)}
\end{equation}
and so, assuming that $H(s)$ has a single stationary point $s_{0}$ in the
half plane $\Re s>\chi\tsub{f}$ we have
\begin{equation}
\chi=H\left(s_{0}\right)
\label{chi_as_Stationary_Value}
\end{equation}

{}From the pole structure of $F$ \Ref{poles} and the fact that $\chi$ is real,
we deduce that $s_{0}$ lies on the real axis, the schematic
form of $H(s)$ being given as in Fig.~\ref{Schematic_Form}. That is, $H(s)$
has a branch point at $s=\chi\tsub{f}$ and
$H(s)\sim s$ as $s\rightarrow\infty$. $s_{0}$ is therefore a local minimum of
$H(s)$ on the real axis viz
\begin{equation}
\chi=\min_{s>\chi\tsub{f}}H(s)
\label{chi_as_Minimum}
\end{equation}

The steepest descents method was first employed by Salsburg, Zwanzig and
Kirkwood \cite{SZK} in their calculation of
equilibrium properties of the Takahashi gas \cite{Takahashi} which, as we
shall in section~\ref{Physical_Interpretation}, has
some similarities with the present model.

The thermodynamics of the model can be determined from $\chi$. For instance,
the fugacity can be rendered as a function of
the temperature, the $d$ electron density $n\equiv N/V$ and the
$f$ electron (well) density $m$ by inverting the relation
\begin{equation}
n=z\frac{\partial\chi}{\partial z}
\label{Density_Equation}
\end{equation}
The Helmholtz free energy is then given by
\begin{equation}
\psi=\mu-\frac{\chi}{n\beta}
\end{equation}
In general, $s_{0}$ cannot be calculated analytically. In the limit
$\beta\rightarrow\infty$ however, the function $H(s)$
simplifies and the ground state energy can be derived.

\section{Ground State Energy}
\label{Ground_State_Energy}
\setcounter{equation}{0}

The ground state energy (per $d$ electron) is given by
\begin{equation}
\varepsilon\tsub{G}=\mu-\frac{p\tsub{G}}{n}
\label{Ground_State1}
\end{equation}
where
\begin{equation}
p\tsub{G}\equiv\lim_{\beta\rightarrow\infty}\frac{\chi}{\beta}
\label{pG}
\end{equation}
denotes the ground state pressure and in \Ref{Ground_State1} $\mu$ is
determined as a function of the $d$ electron density
by inverting the limiting form of \Ref{Density_Equation} viz
\begin{equation}
n=\frac{\partial p\tsub{G}}{\partial\mu}
\label{T=0_Density_Equation}
\end{equation}

To find $p\tsub{G}$ we require the limiting form of $H(s)/\beta$. Now, using
\Ref{chif} we have
\begin{eqnarray}
\chi\tsub{f} & = & \int_{0}^{\infty}\log\left(1+
e^{\beta\left(\mu-x^{2}\right)}\right]dx
\\[0.3em]
& \sim &
\beta\int_{0}^{\sqrt{\mu}}\left(\mu-x^{2}\right)dx
\\[0.3em]
& = & \frac{2}{3}\beta\mu^{3/2}
\label{chif_Large_beta}
\end{eqnarray}

Writing $s=\beta\mu^{3/2}t$ then, we use \Ref{Fcunt} and \Ref{f} to get
\begin{eqnarray}
F(s) & = & \int_{0}^{\infty}
\prod_{k=1}^{\infty}
\left\{1+\exp\left[\beta\mu\left(1-k^{2}/\mu x^{2}\right)\right]\right\}
\exp\left[-\beta\mu t\sqrt{\mu}x\right]dx
\label{First_Step}
\\[0.3em]
& = &
\frac{1}{\sqrt{\mu}}\int_{0}^{\infty}
\prod_{k=1}^{\infty}
\left\{1+\exp\left[\beta\mu\left(1-k^{2}/x^{2}\right)\right]\right\}
\exp\left[-\beta\mu tx\right]dx
\\[0.3em]
& \sim &
\frac{1}{\sqrt{\mu}}\int_{0}^{\infty}
\prod_{k=1}^{\ip{x}}
\exp\left[\beta\mu\left(1-k^{2}/x^{2}\right)\right]
\exp\left[-\beta\mu tx\right]dx
\\[0.3em]
& = &
\frac{1}{\sqrt{\mu}}\int_{0}^{\infty}
e^{\beta\mu g(x,t)}dx
\label{F(s)_Large_beta}
\end{eqnarray}
where
\begin{equation}
g(x,t)\equiv\ip{x}-\frac{\ip{x}(\ip{x}+1/2)(\ip{x}+1)}{3x^{2}}-tx
\label{g(x,t)}
\end{equation}
and we have used the identity
\begin{equation}
\sum_{k=1}^{K}k^{2}=\frac{K(K+1/2)(K+1)}{3}
\label{Sum_Identity}
\end{equation}
Clearly $g(x,t)\sim(2/3-t)x$ as $x\rightarrow\infty$ and hence the integral
\Ref{F(s)_Large_beta} converges if $t>2/3$ as we
would
expect from \Ref{chif_Large_beta} and the requirement that $s>\chi\tsub{f}$.

Evaluating \Ref{F(s)_Large_beta} by Laplace's method we then have
\begin{equation}
F(s)\sim\exp\left[\beta\mu\max_{x\geq 0}g(x,t)\right]
\end{equation}
whence from \Ref{H(s)}, \Ref{chi_as_Minimum} and \Ref{pG} we obtain
\begin{equation}
p\tsub{G}=\min_{t>2/3}h(t)
\label{pG_as_Minimum}
\end{equation}
where
\begin{eqnarray}
h(t) & \equiv &
\lim_{\beta\rightarrow\infty}\frac{H\left(\beta\mu^{3/2}t\right)}{\beta}
\\[0.3em]
& = &
\mu^{3/2}t+m\mu\max_{x\geq 0}g(x,t)
\label{h(t)=}
\end{eqnarray}

$g(x,t)$ is sketched as a function of $x$ for various values of $t$ in
figure~\ref{g(x,t)_Plots}. We see that
${\displaystyle\max_{x\geq 0}}g(x,t)=g(0,t)=0$ for all $t>2/3$. From
\Ref{h(t)=} we therefore have
\begin{equation}
h(t)=\mu^{3/2}t
\end{equation}
and so from \Ref{pG_as_Minimum} we have
\begin{equation}
p\tsub{G}=\frac{2}{3}\mu^{3/2}=
\lim_{\beta\rightarrow\infty}\frac{\chi\tsub{f}}{\beta}
\equiv p\tsub{G}\tsup{(f)}
\label{pGf}
\end{equation}
That is, as $\beta\rightarrow\infty$, $s_{0}\sim\chi\tsub{f}+O(1)$, the large
$\beta$ form of $H(s)$ being as indicated in
Fig.~\ref{Schematic_Form}.

$p\tsub{G}$ therefore equates to $p\tsub{G}\tsup{(f)}$---the ground state
pressure for a free system of $d$ electrons.
Using \Ref{T=0_Density_Equation} we get
\begin{eqnarray}
\mu=n^{2}
\end{eqnarray}
and hence the ground state energy is
\begin{equation}
\varepsilon\tsub{G}=\varepsilon\tsub{G}\tsup{(f)}
\label{eG}
\end{equation}
where $\varepsilon\tsub{G}\tsup{(f)}$ denotes the well known free fermion
ground state energy
\begin{equation}
\varepsilon\tsub{G}\tsup{(f)}=\frac{n^{2}}{3}
\label{eGf}
\end{equation}
The ground state is therefore a state where all of the $f$ particles are
clustered at the ends and all but one of the wells is
unoccupied and of zero width, the $d$ electrons occupying the large well.
In the following section we interpret this result by considering the analogy
between the present model and the Takahashi
gas.

\section{Physical Interpretation and Inclusion of a Takahashi Interaction
Between $f$ Electrons}
\label{Physical_Interpretation}
\setcounter{equation}{0}

In order to physically interpret the above result we recall that, for a
Takahashi gas \cite{Takahashi}, we have a system of $M$
particles at positions $0<y_{1}<\ldots<y_{M}<V$ with a repulsive interaction
\begin{equation}
\sum_{0\leq i<j\leq M+1}\phi\left(y_{j}-y_{j-1}\right)
\end{equation}
where
\begin{equation}
\phi(x)=
\left\{
\begin{array}{ll}
\infty & 0\leq x\leq a
\\[\arrayStretch]
v(x) & a<x<2a
\\[\arrayStretch]
0 & x\geq 2a
\end{array}
\right.
\label{Takahashi_Potential}
\end{equation}
and in \Ref{Takahashi_Potential} $a>0$ is the hard core exclusion distance
and $v(x)>0$.

Because of the hard core exclusion, the interaction only acts between nearest
neighbours and the partition function is
\begin{eqnarray}
Z_{M}
& = &
\int_{0}^{V}
e^{-\beta\phi\left(V-y_{M}\right)}dy_{M}
\int_{0}^{y_{M}}e^{-\beta\phi\left(y_{M}-y_{M-1}\right)}
dy_{M-1}
\nonumber
\\[0.3em]
& & \hspace{15em}\ldots\int_{0}^{y_{2}}
e^{-\beta\phi\left(y_{2}-y_{1}\right)}
e^{-\beta\phi\left(y_{1}\right)}dy_{1}
\label{Takahashi_Partition_Function}
\end{eqnarray}
As mentioned in section~\ref{Continuum_Model}, Laplace convolution and the
method of steepest descents have been
employed \cite{SZK} to solve for the Helmholtz free energy of the Takahashi
gas.

Taking \Ref{Takahashi_Partition_Function} as a starting point, we could
define a model partition function with longer range
forces by relaxing the requirement that $\phi(x)$ vanishes for $x\geq 2a$. As
pointed out however \cite{Lieb_Mattis_Book}
such a partition function is unphysical in that only nearest neighbours
interact and the specific two-body potential form
\Ref{Takahashi_Potential} is required for \Ref{Takahashi_Partition_Function}
to be valid.

We note however that the grand canonical partition function from the present
model
\Ref{Grand_Canonical_Partition_Function} is of the form
\Ref{Takahashi_Partition_Function} if we identify
$\phi(x)$ with $\frac{\log f(x)}{\beta}$.
That is, the nearest neighbour potential between $f$ electrons can be
considered an effective interaction induced by the $d$
electrons. As can be seen from \Ref{f}, this interaction is temperature
dependent, long ranged and attractive.
This sheds some light on the clustering of the $f$ electrons observed in
section \Ref{Ground_State_Energy}. A further
physical interpretation arising from an approximate evaluation of the
thermodynamic potential is given in
appendix~\ref{Low_Temperature}.

Intuitively, from the results mentioned in
section~\ref{Introduction_Falicov_Kimball} for the lattice Falicov-Kimball
model
we might expect that the ground state $f$ electron configuration would be the
analogue of the checkerboard configuration
with the $f$ particles homogeneously spaced along the line. In
appendix~\ref{Equal_Spacing} we derive the ground state
energy from the equally spaced configuration
\begin{equation}
\varepsilon\tsub{G}\tsup{(ES)}=
\frac{m^{3}}{3n}\left(\ip{\frac{n}{m}}+1\right)
\left[
3\left(\ip{\frac{n}{m}}+1\right)
\left(\frac{n}{m}-\ip{\frac{n}{m}}\right)+\ip{\frac{n}{m}}
\left(\ip{\frac{n}{m}}+\frac{1}{2}\right)
\right]
\label{eGES}
\end{equation}
We show explicitly that this lies above the exact ground state energy
\Ref{eGf} which arises from the highly inhomogeneous
$f$ electron configuration mentioned in section~\ref{Ground_State_Energy}.

It is natural then to investigate the effect of including a Takahashi term
\Ref{Takahashi_Potential} (a short range repulsive
potential between $f$ particles) in the model. In this case the barriers
cannot cluster together and the resulting model should
more closely resemble the lattice model where the $f$ electrons cannot
cluster into a very small region.

With the Takahashi term included, the grand canonical partition function has
the form
\Ref{Grand_Canonical_Partition_Function} but with $f$ now defined by
\begin{equation}
f(x)\equiv\prod_{k=1}^{\infty}\left[1+ze^{-\beta k^{2}/x^{2}}\right]
e^{-\beta\phi(x)}
\label{f_Modified}
\end{equation}
The pole structure of $F(s)$ does not change and the thermodynamic potential
is still obtained from \Ref{chi_as_Minimum}.

We restrict our attention to the Tonks case \cite{Tonks} where $v(x)\equiv 0$
in order to calculate the effect that a repulsive
$f$-$f$ interaction has on the ground state. As the hard core exclusion
distance $a$ is increased from $0$ to the extreme
value
$1/m$, the number of allowed barrier configurations is reduced until the only
allowed configuration is the homogeneous
configuration. The ground state energy $\varepsilon\tsub{G}$ should in turn
increase monotonically from the free fermion
value \Ref{eGf} to the equal spacing value \Ref{eGES}. Alternatively, the
ground state pressure $p\tsub{G}$ should
decrease monotonically from the free fermion value \Ref{pGf} to the equal
spacing value
\begin{equation}
p\tsub{G}\tsup{(ES)}=m\mu\ip{\frac{\sqrt{\mu}}{m}}
-\frac{m^{3}}{3}\ip{\frac{\sqrt{\mu}}{m}}
\left(\ip{\frac{\sqrt{\mu}}{m}}+1/2\right)
\left(\ip{\frac{\sqrt{\mu}}{m}}+1\right)
\label{pGES}
\end{equation}
which is derived in appendix~\ref{Equal_Spacing}.

To derive the ground state pressure we repeat steps
\Ref{First_Step}-\Ref{F(s)_Large_beta} with the modified form
\Ref{f_Modified} viz
\begin{eqnarray}
F(s) & = & \int_{0}^{\infty}
\prod_{k=1}^{\infty}
\left\{1+\exp\left[\beta\mu\left(1-k^{2}/\mu x^{2}\right)\right]\right\}
\exp\left[-\beta\mu t\sqrt{\mu}x\right]
e^{-\beta\phi(x)}dx
\\[0.3em]
& = & \int_{a}^{\infty}
\prod_{k=1}^{\infty}
\left\{1+\exp\left[\beta\mu\left(1-k^{2}/\mu x^{2}\right)\right]\right\}
\exp\left[-\beta\mu t\sqrt{\mu}x\right]dx
\\[0.3em]
& = &
\frac{1}{\sqrt{\mu}}\int_{a\sqrt{\mu}}^{\infty}
\prod_{k=1}^{\infty}
\left\{1+\exp\left[\beta\mu\left(1-k^{2}/x^{2}\right)\right]\right\}
\exp\left[-\beta\mu tx\right]dx
\\[0.3em]
& \sim &
\frac{1}{\sqrt{\mu}}\int_{a\sqrt{\mu}}^{\infty}
e^{\beta\mu g(x,t)}dx
\end{eqnarray}

Applying Laplace's method once more we have
\begin{equation}
F(s)\sim\exp\left[\beta\mu\max_{x\geq a\sqrt{\mu}}g(x,t)\right]
\end{equation}
and we again have \Ref{pG_as_Minimum} but with \Ref{h(t)=} generalised to
\begin{equation}
h(t)=\mu^{3/2}t+m\mu\max_{x\geq a\sqrt{\mu}}g(x,t)
\end{equation}

Clearly $h(t)$ and therefore $p\tsub{G}$ is a monotonically decreasing
function of $a$. As $a$ approaches $1/m$ we find
that
\begin{eqnarray}
\max_{x\geq a\sqrt{\mu}}g\left(x,t_{0}\right)
& = &
g\left(a\sqrt{\mu},t_{0}\right)
\\[0.3em]
& \rightarrow &
g\left(\frac{\sqrt{\mu}}{m},t_{0}\right)
\\[0.3em]
& = &
\ip{\frac{\sqrt{\mu}}{m}}
-\frac{m^{2}}{3\mu}\ip{\frac{\sqrt{\mu}}{m}}
\left(\ip{\frac{\sqrt{\mu}}{m}}+1/2\right)
\left(\ip{\frac{\sqrt{\mu}}{m}}+1\right)
-\frac{\sqrt{\mu}t_{0}}{m}
\end{eqnarray}
where $t_{0}\equiv{\displaystyle\lim_{\beta\rightarrow\infty}}
s_{0}/\beta\mu^{3/2}$ is the minimizing value in \Ref{pG_as_Minimum}. As
expected, it follows from
\Ref{pGES} and \Ref{pG_as_Minimum} then that
$p\tsub{G}\rightarrow p\tsub{G}\tsup{(ES)}$ as $a\rightarrow 1/m$.

Numerical investigations of $p\tsub{G}$ as a function of $a$ reveal that the
decrease from $p\tsub{G}\tsup{(f)}$ to
$p\tsub{G}\tsup{(ES)}$ is not smooth. We illustrate this for the case where
$\mu=5$ and $m=2$ in Fig.~\ref{h(t)_Plots}
where we plot $h(t)$ versus $t$ for various values of $a$ and
Fig.~\ref{pG(a)} where we plot where we plot $p\tsub{G}$
as a function of $a$. We see that there is a critical value of the exclusion
$a\tsub{c}$ above which $p\tsub{G}$ equates to
$p\tsub{G}\tsup{(ES)}$. We are naturally lead to the following
characterization of the ground state $f$ electron
configuration.

For $a=0$ the $f$ particles lie in the highly inhomogeneous configuration
where all wells have zero width except one which
has width $V$. For $0<a<a\tsub{c}$ all wells must be at least as wide as $a$
but inhomogeneity in the $f$ configuration
persists with one well wider than the others. For $a\tsub{c}\leq a\leq 1/m$
the ground state configuration is the homogeneous
or equal spaced configuration, each well having width $1/m$ even though
inhomogeneous configurations are allowed for
$a\tsub{c}\leq a<1/m$.

We note from Fig.~\ref{pG(a)} that $p\tsub{G}$ may in general have further
singularities in the region $0<a<a\tsub{c}$.
These singularities relate not to the well configuration but to the
distribution (or filling) of the $d$ electrons in the wells. For
sufficiently small $a$, the large well has width $V-Ma$ (all the small wells
have the minimal width $a$) and all of the $d$
electrons lie in the large well. That is, the ground state pressure is that
of a free fermion system with a reduced volume. It is
easy to show that in this case the ground state pressure is $(1-
ma)p\tsub{G}\tsup{(f)}$. As $a$ is increased towards
$a\tsub{c}$, critical values occur at which it becomes energetically
favourable to fill the smaller wells. In the specific case
depicted, there is a single critical value of the exclusion
$\tilde{a}\tsub{c}\in\left(0,a\tsub{c}\right)$ such that $p\tsub{G}$
equates to $(1-ma)p\tsub{G}\tsup{(f)}$ for $0<a\leq\tilde{a}\tsub{c}$ and
filling of the smaller wells occurs for
$\tilde{a}\tsub{c}<a<a\tsub{c}$.

\section{Discussion}
\setcounter{equation}{0}

In this chapter we have studied the thermodynamics of a one dimensional
continuum analogue of the Falicov-Kimball
model in the limit of strong correlation. As mentioned in the introduction,
in terms of mathematical and physical complexity,
this model is a long way removed from a lattice model with
finite strength interactions in higher dimensions . As shown in
section~\ref{Physical_Interpretation} however, the exact solution does afford
a neat physical interpretation and the addition of
a repulsive $f$-$f$ interaction enriches the model leading to a ground state
analogous to that observed in lattice models if the
repulsion is sufficiently strong. Finally,
we indicate some directions for further work.

The calculation of the ground state $f$ electron configuration and $d$
electron distribution in the case where the $f$-$f$
interaction is of the Tonks form could be taken further and made more
precise. Also, a study of the ground state when
$v(x)$ is non-zero may prove interesting because in this case the Takahashi
repulsion will contribute directly to the ground
state
energy and not simply through the reduction of allowed $f$ configurations. It
should be possible to generalise to the present
model a method used
by Salsburg, Zwanzig and Kirkwood in the calculation of correlation functions
for the Takahashi gas \cite{SZK}. The
calculation of the $f$-$f$ correlation function, especially at zero
temperature may be of some interest. It
may be of interest to generalise the model to higher dimensions $d$ where the
barriers are impenetrable $d-1$ dimensional
hyperplanes. Finally, it might be interesting to investigate this model in
the case where the $d$ particles have a different
energy spectrum, different statistics, or both.

The author was supported by an Australian Postgraduate Research Award
throughout the course of this work. The author acknowledges the encouragement
of his Ph. D. supervisor, Professor Colin Thompson and is
 thankful to Dr. Bill Wood for some useful discussions.

\section*{Appendix}
\appendix

\section{A Representation for $f$}
\label{Infinite_Product}
\setcounter{equation}{0}

In this appendix we derive the representation \Ref{Large_x_Representation}
for the function $f$. Now from \Ref{f} we
have
\begin{equation}
\log f(x)=\frac{1}{2}\sum_{k=-\infty}^{\infty}{\cal F}(k/x)
-\frac{1}{2}\log(1+z)
\label{logf=}
\end{equation}
where
\begin{equation}
{\cal F}(w)\equiv\log\left(1+ze^{-\beta w^{2}}\right)
\end{equation}

${\cal F}(w)$ is absolutely integrable so the Poisson summation formula can
be applied to
\Ref{logf=} viz
\begin{eqnarray}
\log f(x) & = &
\frac{x}{2}\sum_{k=-\infty}^{\infty}\tilde{{\cal F}}(2\pi kx)
-\frac{1}{2}\log(1+z)
\\[0.3em]
& = &
x\chi\tsub{f}+x\sum_{k=1}^{\infty}\tilde{{\cal F}}(2\pi kx)
-\frac{1}{2}\log(1+z)
\label{logf(x)=}
\end{eqnarray}
where
\begin{equation}
\tilde{{\cal F}}(y)\equiv\int_{-\infty}^{\infty}
{\cal F}(w)e^{iwy}dw
\end{equation}
denotes the Fourier transform of ${\cal F}$ and we have noted using
\Ref{chif} that
$\tilde{{\cal F}}(0)=2\chi\tsub{f}$.

On integrating by parts we find that
\begin{equation}
\tilde{{\cal F}}(y)=-\frac{1}{iy}\int_{-\infty}^{\infty}
{\cal F}'(w)e^{iwy}dw
\label{tiF(y)=}
\end{equation}
Now
\begin{equation}
-{\cal F}'(w)=
\frac{2\beta wze^{-\beta w^{2}}}{1+ze^{-\beta w^{2}}}
\end{equation}
so, above the real axis $-{\cal F}'(w)$ has simple poles at
$w=w_{l}^{\pm}=\pm u_{l}+iv_{l}$ $l=1,2,3,\ldots$ where $u_{l}$ and $v_{l}$
are the positive, real solutions of
\begin{equation}
\beta\left(w_{l}^{\pm}\right)^{2}=\beta\mu\pm(2l-1)\pi i
\end{equation}
That is, the intersection of the two hyperboli
\begin{eqnarray}
u_{l}v_{l} & = & \frac{(2l-1)\pi}{2\beta}
\\[0.3em]
u_{l}^{2}-v_{l}^{2} & = & \mu
\end{eqnarray}
The explicit values of $u_{l}$ and $v_{l}$ are given by \Ref{uk} and
\Ref{vk}. The residue of $-{\cal F}'(w)e^{iwy}$ at
$w=w_{l}^{\pm}$ is $-e^{iw_{l}^{\pm}y}$. Assuming $y>0$, using \Ref{tiF(y)=}
and closing the contour above the
real axis\footnoteStretch\footnote{It is easily seen that the integral on the
closing contour approaches zero as long as the
closing contour is chosen to bisect the poles so that the denominator of
$-{\cal F}'(w)$ is bounded away from
$0$.}\unStretch~we therefore have
\begin{equation}
\tilde{{\cal F}}(y)=
-\frac{2\pi}{y}\sum_{l=1}^{\infty}\left(e^{iw_{l}^{+}y}+
e^{iw_{l}^{-}y}\right)
\label{F(y)=}
\end{equation}

Combining \Ref{logf(x)=} with \Ref{F(y)=} we arrive at
\begin{equation}
\log f(x)=
x\chi\tsub{f}-\frac{1}{2}\log(1+z)-\sum_{k=1}^{\infty}\frac{1}{k}
\sum_{l=1}^{\infty}\left(e^{2\pi ikw_{l}^{+}x}+
e^{2\pi ikw_{l}^{-}x}\right)
\end{equation}
Using the Taylor expansion
\begin{equation}
\sum_{k=1}^{\infty}\frac{r^{k}}{k}=-\log(1-r)
\end{equation}
we then have
\begin{equation}
\log f(x)=x\chi\tsub{f}-\frac{1}{2}\log(1+z)
+\sum_{l=1}^{\infty}\log\left(
1-e^{2\pi iw_{l}^{+}x}\right)
\left(1-e^{2\pi iw_{l}^{-}x}\right)
\end{equation}
from which the required representation \Ref{Large_x_Representation} follows.

\section{Approximate Low Temperature Calculation of the Thermodynamic
Potential and Further Physical Interpretation}
\label{Low_Temperature}
\setcounter{equation}{0}

In this appendix we give an approximate direct evaluation of the
thermodynamic potential which should be reasonable at very
low temperatures and which admits a neat physical interpretation.

Now using \Ref{Large_x_Representation} and \Ref{Fcunt} we can write
\begin{equation}
F(s)=\frac{1}{\sqrt{1+z}}\left[\frac{1}{s-\chi\tsub{f}}+G(s)\right]
\label{F=}
\end{equation}
where
\begin{equation}
G(s)\equiv\int_{0}^{\infty}
e^{(\chi\tsub{f}-s)x}\left[\prod_{k=1}^{\infty}
\left|1-e^{2\pi\left(-v_{k}
+iu_{k}\right)x}\right|^{2}-1\right]dx
\label{G(s)}
\end{equation}
is analytic for $\Re s>2\pi v_{1}$.

The right hand side of \Ref{Q===FK} is dominated by the pole at
$s=\chi\tsub{f}$ for
large $M$.
The terms arising from the
other singularities of $F$ are exponentially small.
Using \Ref{F=} we write
\begin{equation}
(F(s))^{M+1}e^{sV}=
\frac{e^{\chi\tsub{f}V}}
{(1+z)^{\frac{M+1}{2}}}\sum_{j=0}^{M+1}
{M+1 \choose j}
\frac{\left[G(s)\right]^{M+1-j}}{(s-\chi\tsub{f})^{j}}
\sum_{l=0}^{\infty}\frac{(s-\chi\tsub{f})^{l}V^{l}}
{l!}
\label{residue1}
\end{equation}

{}From \Ref{vk} and \Ref{poles} we see that the poles of $F(s)$ accumulate on
the line $\Re s=\chi\tsub{f}$ as
$\beta\rightarrow\infty$. We therefore make the approximation $G(s)\approx
G\left(\chi\tsub{f}\right)$ and
so from \Ref{residue1} we approximate the residue of $(F(s))^{M+1}e^{sV}$ at
$s=\chi\tsub{f}$ by
\begin{equation}
\frac{e^{\chi\tsub{f}V}}{(1+z)^{\frac{M+1}{2}}}
\sum_{j=1}^{M+1}
{M+1 \choose j}
\frac{\left[G\left(\chi\tsub{f}\right)\right]^{M+1-j}V^{j-1}}{(j-1)!}
\end{equation}
In this approximation then the dominant contribution to the partition
function is
\begin{eqnarray}
{\cal Q}(V) & \approx &
\frac{e^{\chi\tsub{f}V}}{V(1+z)^{\frac{M+1}{2}}}
\sum_{j=1}^{M+1}
{M+1 \choose j}
\frac{\left[G\left(\chi\tsub{f}\right)\right]^{M+1-j}V^{j}}{(j-1)!}
\label{Qapprox}
\end{eqnarray}

Making use of the identity
\begin{equation}
\frac{1}{(j-1)!}=\frac{1}{2\pi i}\int_{c-i\infty}^{c+i\infty}
\frac{e^{s}}{s^{j}}\;ds\hspace{2em}c>0
\end{equation}
we have
\begin{eqnarray}
& &
\sum_{j=1}^{M+1}
{M+1 \choose j}
\frac{\left[G\left(\chi\tsub{f}\right)\right]^{M+1-j}V^{j}}{(j-1)!}
\nonumber
\\[0.3em]
& = &
\frac{1}{2\pi i}\int_{c-i\infty}^{c+i\infty}
ds\;e^{s}
\sum_{j=1}^{M+1}
{M+1 \choose j}
\left[G\left(\chi\tsub{f}\right)\right]^{M+1-j}
\left(\frac{V}{s}\right)^{j}
\\[0.3em]
& = &
\frac{1}{2\pi i}\int_{c-i\infty}^{c+i\infty}
e^{s}
\left\{\left[\frac{V}{s}+G\left(\chi\tsub{f}\right)\right]^{M+1}
-\left[G\left(\chi\tsub{f}\right)\right]^{M+1}\right\}ds
\\[0.3em]
& = &
\frac{V^{M+1}}{2\pi i}
\int_{c-i\infty}^{c+i\infty}
\frac{e^{s}}{s^{M+1}}\left\{
\left[1+\frac{G\left(\chi\tsub{f}\right)s}{V}\right]^{mV}
-\left[\frac{G\left(\chi\tsub{f}\right)s}{V}\right]^{mV}
\right\}ds
\nonumber
\\[0.3em]
& \approx &
\frac{V^{M+1}}{2\pi i}
\int_{c-i\infty}^{c+i\infty}
\frac{ds}{s^{M+1}}\exp
\left[s\left(1+mG\left(\chi\tsub{f}\right)\right)\right]
\\[0.3em]
& = &
\frac{V^{M+1}}{M!}
\left[1+mG\left(\chi\tsub{f}\right)\right]^{M}
\label{sumapprox}
\end{eqnarray}
where we have used the standard limit
\begin{equation}
\lim_{V\rightarrow\infty}\left[1+\frac{a}{V}\right]^{V}=e^{a}
\end{equation}

Combining \Ref{Qapprox} and \Ref{sumapprox} we then have
\begin{equation}
{\cal Q}(V)\approx
\frac{V^{M+1}}{M!}\frac{e^{\chi\tsub{f}V}}{(1+z)^{\frac{M+1}{2}}}
\left[1+mG\left(\chi\tsub{f}\right)\right]^{M}
\end{equation}
Taking the thermodynamic limit ($V\rightarrow\infty$ with $m=(M+1)/V$ fixed)
we obtain the approximate thermodynamic
potential
\begin{equation}
\chi\approx
m-m\log m+\chi\tsub{f}
+m\log\left[\frac{1+mG\left(\chi\tsub{f}\right)}{\sqrt{1+z}}\right]
\label{Approximate_Expression}
\end{equation}

The above expression has a simple interpretation. The first term is an
entropy term arising from the barrier configurations.
The second term is the free fermion term arising from states where all wells
have zero width except one with width $V$. The
last term arises from states where a macroscopic number of wells have
non-zero width.

Finally, from \Ref{G(s)} and \Ref{Large_x_Representation} we have
\begin{equation}
G\left(\chi\tsub{f}\right)=\int_{0}^{\infty}
\left[
\prod_{k=1}^{\infty}
\left|1-e^{2\pi\left(-v_{k}+iu_{k}\right)x}\right|^{2}-1
\right]dx
\end{equation}
Using methods from section~\ref{Ground_State_Energy} it can be shown that the
last term in
\Ref{Approximate_Expression} does not contribute to the ground state pressure
and so we recover the result \Ref{eG}
within
this approximation.

\section{Ground State Energy and Pressure for the Homogeneous $f$ Electron
Configuration}
\label{Equal_Spacing}
\setcounter{equation}{0}

In this appendix we calculate some ground state quantities for the
homogeneous or equal spaced $f$ configuration. That is,
the position of the $j$th barrier is
\begin{equation}
y_{j}=j\frac{V}{M+1}=\frac{j}{m}\hspace{2em}j=1,\ldots,M
\label{Homogeneous_Configuration}
\end{equation}
In this case the $M+1$ wells all have width $1/m$ and so the energies
\Ref{Ejk} reduce to
\begin{equation}
E_{jk}=m^{2}k^{2}
\label{EjkES}
\end{equation}

For $N$ $d$ electrons, the lowest energy is achieved by filling the lowest
energy states of the wells with as close to equal
filling as possible. We write
\begin{equation}
N=(q+r)(M+1)
\end{equation}
where
\begin{equation}
q\equiv\ip{\frac{N}{M+1}}=\ip{\frac{n}{m}}
\label{q}
\end{equation}
is the integral part of $n/m$ and
\begin{equation}
r\equiv\frac{N}{M+1}-\ip{\frac{N}{M+1}}
=\frac{n}{m}-\ip{\frac{n}{m}}\in[0,1)
\label{r}
\end{equation}
The first $q$ levels are filled in each well and the $(q+1)$th level is
filled in $(M+1)r$ of the wells. Using \Ref{EjkES} then,
the ground state energy per $d$ particle is
\begin{eqnarray}
\varepsilon\tsub{G}\tsup{(ES)}
& = &
\frac{1}{N}\left[(M+1)\sum_{k=1}^{q}m^{2}k^{2}+(M+1)rm^{2}(q+1)^{2}
\right]
\\[0.3em]
& = &
\frac{m^{3}}{n}\left[\frac{q(q+1/2)(q+1)}{3}+r(q+1)^{2}\right]
\end{eqnarray}
where we have made use of \Ref{Sum_Identity}. Using \Ref{q}, \Ref{r} and a
little manipulation, we then arrive at
\Ref{eGES}.

We next show explicitly that the exact ground state energy \Ref{eG} (from the
highly inhomogeneous configuration) lies
below the ground state energy from the homogeneous configuration. Using
\Ref{eG}, \Ref{eGf}, \Ref{eGES}, \Ref{q} and
\Ref{r} we have
\begin{eqnarray}
\varepsilon\tsub{G}\tsup{(ES)}
& = &
\frac{m^{3}}{3n}(q+1)\left\{
3(q+1)r+q(q+1/2)
\right\}
\\[0.3em]
& = &
\frac{m^{3}}{3n}(q+1)\left\{
q^{2}+(3r+1/2)q+3r
\right\}
\\[0.3em]
& = &
\frac{m^{3}}{3n}(q+1)\left\{
\left(q+\frac{(3r+1/2)}{2}\right)^{2}+3r-
\frac{(3r+1/2)^{2}}{4}
\right\}
\\[0.3em]
& = &
\frac{m^{3}}{3n}(q+1)\left\{
\left(q+r+\frac{r+1}{2}\right)^{2}+3r-\frac{9r^{2}}{4}
-\frac{3r}{4}-\frac{1}{16}
\right\}
\\[0.3em]
& = &
\frac{m^{3}}{3n}(q+1)\left\{
(q+r)^{2}+(r+1)(q+r)+\left(\frac{r+1}{2}\right)^{2}
+3r
\right.
\nonumber
\\[0.3em]
& &
\hspace{16em}
\left.
-\;\frac{r(9r+3)}{4}-\frac{1}{16}
\right\}
\\[0.3em]
& \geq &
\frac{m^{3}}{3n}(q+1)\left\{
(q+r)^{2}+\frac{1}{4}+3r-\frac{12r}{4}-\frac{1}{16}
\right\}
\\[0.3em]
& \geq &
\frac{m^{3}}{3n}(q+r)^{3}
\\[0.3em]
& = &
\frac{n^{2}}{3}
\\[0.3em]
& = &
\varepsilon\tsub{G}\tsup{(f)}
\\[0.3em]
& = &
\varepsilon\tsub{G}
\end{eqnarray}
as required.

Finally, we calculate the ground state pressure from the homogeneous
configuration in terms of $m$ and $\mu$. To do so
we find the grand canonical partition function for the specific barrier
configuration \Ref{Homogeneous_Configuration}. The
canonical partition function is
\begin{equation}
Z_{MN}\tsup{(ES)}=
\sum_{j=1}^{M+1}\sum_{k=1}^{\infty}\sum_{\left\{n_{jk}\right\}}'
e^{-\beta E_{jk}n_{jk}}
\end{equation}
and, using \Ref{EjkES}, the associated grand canonical partition function is
\begin{eqnarray}
{\cal Q}\tsup{(ES)}(V)
& = &
\sum_{N=0}^{\infty}z^{N}Z_{MN}\tsup{(ES)}
\\[0.3em]
& = &
\prod_{j=1}^{M+1}\prod_{k=1}^{\infty}
\left(1+ze^{-\beta m^{2}k^{2}}\right)
\\[0.3em]
& = &
\left[\prod_{k=1}^{\infty}
\left(1+ze^{-\beta m^{2}k^{2}}\right)
\right]^{M+1}
\end{eqnarray}

The thermodynamic potential is
\begin{eqnarray}
\chi\tsup{(ES)}(V)
& = &
\lim_{V\rightarrow\infty}
\frac{\log{\cal Q}\tsup{(ES)}}{V}
\\[0.3em]
& = &
m\sum_{k=1}^{\infty}\log
\left\{1+\exp\left[\beta\mu\left(1-m^{2}k^{2}/\mu\right)
\right]\right\}
\\[0.3em]
& \sim &
m\beta\mu\sum_{k=1}^{\ip{\sqrt{\mu}/m}}
\left[1-\frac{m^{2}k^{2}}{\mu}\right]\hspace{2em}
\mbox{(for large $\beta$)}
\\[0.3em]
& = &
m\beta\mu\ip{\frac{\sqrt{\mu}}{m}}
-\frac{m^{3}\beta}{3}\ip{\frac{\sqrt{\mu}}{m}}
\left(\ip{\frac{\sqrt{\mu}}{m}}+1/2\right)
\left(\ip{\frac{\sqrt{\mu}}{m}}+1\right)
\end{eqnarray}
so the ground state pressure $p\tsub{G}\tsup{(ES)}
\equiv{\displaystyle\lim_{\beta\rightarrow\infty}}\frac{\chi\tsup{(ES)}}
{\beta}$ is given by \Ref{pGES}.


\vfill
\eject

\section*{List of Figures}

\noindent 1. The schematic form of $H(s)$ on the real axis (solid line)
and the large $\beta$ form (dot-dashed line).

\noindent 2. $g(x,t)$ as a function of $x$ for various values of $t$.

\noindent 3. $h(t)$ as a function of $t$ for various values of
the exclusion distance $a$ in the case where $\mu=5$ and $m=2$.

\noindent 4. Ground state pressure $p\tsub{G}$ (solid line) as
a function of the exclusion distance $a$ in the case where
$\mu=5$ and $m=2$. $p\tsub{G}$ equates to $(1-ma)p\tsub{G}\tsup{(f)}$
(dot-dashed line)---the ground state pressure
from a free fermion system with restricted volume
$V-Ma$---in the region $0<a<\tilde{a}\tsub{c}$.

\vfill
\eject

\figureStretch
\begin{figure}[p]
\caption{\figsz The schematic form of $H(s)$ on the real axis (solid line)
and the large $\beta$ form (dot-dashed line).}
\label{Schematic_Form}
\end{figure}
\unStretch

\figureStretch
\begin{figure}[p]
\caption{\figsz $g(x,t)$ as a function of $x$ for various values of $t$.}
\label{g(x,t)_Plots}
\end{figure}
\unStretch

\figureStretch
\begin{figure}[p]
\caption{\figsz $h(t)$ as a function of $t$ for various values of the
exclusion distance $a$ in the case where $\mu=5$ and
$m=2$.}
\label{h(t)_Plots}
\end{figure}
\unStretch

\figureStretch
\begin{figure}[h]
\caption{\figsz Ground state pressure $p\tsub{G}$ (solid line) as a function
of the exclusion distance $a$ in the case where
$\mu=5$ and $m=2$. $p\tsub{G}$ equates to $(1-ma)p\tsub{G}\tsup{(f)}$
(dot-dashed line)---the ground state pressure
from a free fermion system with restricted volume $V-Ma$---in the region
$0<a<\tilde{a}\tsub{c}$.}
\label{pG(a)}
\end{figure}
\unStretch

\end{document}